\documentclass[graybox]{svmult}

\usepackage{mathptmx}       
\usepackage{helvet}         
\usepackage{courier}        
\usepackage{type1cm}        
\usepackage{makeidx}         
\usepackage{graphicx}        
\usepackage{multicol}        
\usepackage[bottom]{footmisc}

\makeindex             
\usepackage{graphicx}
\usepackage{amsmath}
\usepackage{color}
\usepackage[normalem]{ulem}
\usepackage{multirow}

\DeclareRobustCommand\bblash{\btt{\@backslashchar}} \makeatother
\begin{document}

\title*{Understanding General Relativity after 100 years: A matter of perspective}
\author{Naresh Dadhich}

\institute{Naresh Dadhich \at Center for Theoretical Physics, Jamia Millia Islamia, New Delhi 110 025, India\\ 
and\\
Inter-University Center for Astronomy and Astrophysics, Post Bag 4, Pune 411 007, India\\
\email{nkd@iucaa.in}}

\maketitle

\abstract*{This is the centenary year of general relativity, it is therefore natural to reflect on what perspective we have evolved in 100 years. I wish to share here a novel perspective, and the insights and directions that ensue from it.}

\abstract{This is the centenary year of general relativity, it is therefore natural to reflect on what perspective we have evolved in 100 years. I wish to share here a novel perspective, and the insights and directions that ensue from it.}

\section{Prologue}

It is the contradiction between observed phenomena and the prevalent theory that drives  search for a new theory. Then it takes few to several years of intense activity of  tentative guesses, effective and workable proposals, and slowly a new understanding evolves inch by inch that finally leads to new theory. This is indeed the case for all physical theories. Take the case of special relativity (SR). Electromagnetic theory when it was ultimately completed by Maxwell by synthesizing Coulomb, Ampere and Faraday, and introducing the ingenious displacement current in 1875, it predicted an invariant velocity of light. The famous Michelson-Morley experiment verified the prediction with great accuracy leaving no room for any suspicion or doubt. The universally constant  velocity obviously conflicted with the Newtonian mechanics. Then the search began which culminated in 1905 after 30 years in Einstein's discovery of special relativity (SR). \\

In the journey to special relativity there were contributions from severe people, most notably of Poincare and Lorentz who had it all but for one bold statement that velocity of light is universally constant. Then came a young man of 26, and simply did, what Poincare and Lorentz hesitated to do, and walked away with the credit of discovering one of the fundamental theories of physics. He said that the velocity of light was constant and its incorporation in mechanics naturally led to special relativity. After 30 years of probing, the atmosphere was sufficiently charged for someone to take the crucial bold step to pick up lowly hanging SR. It was strange that Poincare and Lorentz, who had explored various  properties of SR, failed to take the critical step. It was perhaps their great scientific reputation that came in the way. If it did not turn out right, their hard earned reputation over a life time would go down the drain and the whole world would laugh at them. It was this hesitation that costed them 
dearly in losing their rightful claim on discovery of SR. On the other hand Einstein had neither much reputation nor even an academic job to worry about. He had nothing at stake as he was a clerk in a patent office. If it did not come out right, nothing much would have been lost. \\

Like all other discoveries it was a situational discovery. If it were not him, someone else would have done it in a year or two. Had he discovered only SR, he would have been one among many great scientists, but not in a different league altogether. For that he had to do something very special which none else could have done. By that I mean that when atmosphere is sufficiently charged and ripe for a new theory to sprout, it is a matter of chance, who happened to take the critical last step. So far as SR was concerned, Einstein was really lucky. \\

Following SR, the real action at that time was in understanding atomic structure by building a new theory of quantum mechanics. He did make a pretty interesting little  contribution in that which was good enough to win him the Nobel prize. But then he totally withdrew from the action and devoted 10 long years for completion of the principle of relativity, from special to general. In the process, he arrived at a new relativistic theory of gravity -- general relativity (GR). True, there was no observation or experiment that asked for anything beyond the Newtonian theory at that time. He was therefore not driven by contradiction with experiment but was entirely propelled by the principle. That is why GR was born as a whole and also much ahead of its time. This is something none else could have discovered. \\

To put it all in perspective, had it not been for him, nobody would have asked for a new theory of gravity until quasars were discovered in mid 1960s. This is what puts him in a class of his own. And so is GR as well because it was born out of a principle without any bearing on observation and experiment, whatsoever. More importantly it makes demand on spacetime, which no other force makes, that it has to curve to describe its dynamics. Above all, not only it still stand tall and firm after 100 years, but its centenary is in fact being celebrated with the detection of gravitational waves which were also predicted by Einstein 100 years back! This is indeed a discovery of the same proportion as that of the electromagnetic waves, and hence it is one of the greatest of all times. It is time to salute with utmost reverence and admiration both the theory as well as its creator. 

\section{Introduction}

General relativity is in many ways unique and different from all other physical theories. The first and foremost among them is the fact that, unlike all other forces, relativistic gravitational law is not prescribed but instead it is dictated by spacetime geometry itself. It naturally arises from inhomogeneity of spacetime and that is why it is universal -- links to everything that physically exists. Presence of any force makes spacetime inhomogeneous for particles to which the force links but not for others. For instance, presence of electric field makes spacetime inhomogeneous for electrically charged particles while for neutral particles it remains homogeneous. By universal  force we mean a force that links to everything that physically exists irrespective of particle parameters like mass, charge and spin. Since relativistic gravity is universal and hence it can only be described by spacetime geometry. Thus unlike Newton, Einstein had no freedom to prescribe a relativistic gravitational law because it is 
entirely governed by spacetime itself which does not obey anyone's dictate or prescription. Since relativistic gravity encompasses Newtonian gravity, it is remarkable that now Newton's inverse square law simply follows from spacetime geometry without any external prescription. \\

Note that spacetime is a universal entity as it is the same for all and equally shared by all and so is the universal force. Hence the two respond to each-other leaving no room for any external intervention. By simply appealing to inhomogeneity of spacetime curvature, we will derive an equation of motion for universal force which would be nothing other than Einsteinian gravity. It is remarkable that we make no reference to gravity at all yet  spacetime curvature yields gravitational equation. This happens 
because both \textit{spacetime and Einstein gravity are universal} \cite{measure-grav}. A general principle that emerges is that \textit{all universal things respond to each other and they must therefore be related}. \\

The equation of motion that emerges from Riemann curvature is non-linear involving square of first derivative of metric. It indicates that gravity is self interactive. As a matter of fact it is the universal character that demands energy in any form must gravitate. Since gravitational field like any other field has energy, it must hence also gravitate -- self interact.  Isn't it wonderful that spacetime curvature automatically incorporates this feature through nonlinearity of Riemann tensor? The important aspect of self interaction is that it gravitates without changing the Newtonian inverse square law. This is rather strange because self interaction would, in the classical framework, have asked for $\nabla^2\Phi = 1/2 \Phi'^2$ which would have disturbed the inverse square law. The situation is exactly as it is for photon (light) to feel gravity without having to change its velocity. Within classical framework it is impossible to accommodate these contradictory demands. \\

The answer could however be in the enlargement of framework in which gravity curves space and photon freely floats on it without having to change its velocity. What should curve space and the obvious answer is gravitational field energy which is not supposed to contribute to acceleration, $\nabla\Phi$. Thus gravity self interacts via space curvature and that also facilitates photon's interaction with gravity \cite{ein-new}. These are the two new aspects of Einstein gravity which wonderfully take care of each-other leaving Newtonian inverse square law intact. Einstein is therefore Newton with space curved \cite{ein-new}. Since space and time are bound together in spacetime through universal light velocity, and hence spacetime must be curved. This is how spacetime curvature enters in description of the universal force -- Einstein gravity. In this way the self interaction gets automatically incorporated in Riemann curvature and is reflected through occurrence of square of first derivative of metric. \\

General relativity (GR) is undoubtedly the most elegant and beautiful theory and it is for nothing that Paul Dirac termed it as the greatest feat of human thought! \\

In what follows we would further explore its elegance and richness of structure and form in relation to what new insights and understanding we have gained in past 100 years, and marvel on new questions and directions that ensue. 

\section{At the very beginning} 

Let us begin by characterizing free state of space and time in absence of all forces. Space is homogeneous and isotropic and time is homogeneous. As space is homogeneous, one can freely interchange $x$ and $y$. Since time is also homogeneous which means both space and time are homogeneous, and hence $x$ and $t$ should also be similarly interchangeable. But they are not of the same dimension. Never mind, homogeneity of space and time is a general property which must always be respected and it demands their interchange. One has therefore to bring $x$ and $t$ to the same dimension by demanding existence of a universal invariant velocity $c$ so that $x$ and $ct$ could be interchanged \cite{measure-grav}. Thus it is homogeneity of space and time that demands existence of a universally constant velocity without reference to anything else. It is identified through Maxwell's electrodynamics with velocity of light. Thus space and time get bound together into spacetime through the universal  velocity of light. It thus 
arises in a natural way as a constant of spacetime structure independent of anything else. \\ 

Spacetime free of all dynamics and forces is therefore homogeneous (space and time being homogeneous and space being isotropic). The next question that arises is, what is its geometry? As spacetime is homogeneous so should be its geometry. Geometry is defined by Riemann curvature and so it should be homogeneous which means it should be covariantly constant;i.e. $R_{abcd_;e}=0$. The Riemann curvature should therefore be written in terms something which is constant for covariant derivative. That is the metric tensor and hence we write 
\begin{equation}
 R_{abcd} = \Lambda(g_{ac}g_{bd}-g_{ad}g_{bc}).
\end{equation}
A homogeneous spacetime free of all forces is thus a spacetime of constant curvature, $\Lambda$ and not necessarily Minkowskian of zero curvature. It is a maximally symmetric spacetime and that is what is required for absence of all dynamics. The important point to note is that Minkowski is not dictated by homogeneity of spacetime but it is rather an external imposition by setting $\Lambda=0$. Of course one is free to choose $\Lambda$ zero but then it has to be justified on physical grounds. In classical physics, force free state is characterized by constant potential or constant velocity while for the Einstein gravity it is done by constant curvature. The important point to realize is that constant curvature means no dynamics -- it is on the same footing as constant potential in classical physics. This is because it is Riemann curvature which is the basic element for description of the Einstein gravity and hence it is this, like potential for classical physics, that should have freedom of addition of a 
constant. Force free homogeneous spacetime is in general described by maximally symmetric dS/AdS and not necessarily by flat Minkowski. \\

The point we wish to emphasize is that $\Lambda$ arises naturally as a constant of spacetime structure on the same footing as $c$ without reference to any physical force or phenomenon. These two are pure constants of spacetime structure itself and arise as the  characteristics of \emph{force free} state. They are therefore the most fundamental  constants of Nature. No other constant can claim this degree of fundamentalness simply because none else arises directly from spacetime structure itself. \\

The next question that arises is, what happens when spacetime is not homogeneous?  Obviously it should indicate presence of force which makes spacetime inhomogeneous for all particles irrespective of their mass, charge or any other attributes. This force should therefore be universal meaning it links to everything that physically exists. Everything that physically exists must have energy-momentum -- a universal attribute/charge, and hence this force must link to the universal charge -- energy-momentum. How do we then determine its dynamics? Since presence of this universal force makes spacetime curvature inhomogeneous, hence its dynamics cannot be prescribed but has to follow from the curvature itself. What is it that we can do to Riemann curvature to get to an equation of motion for the universal force? The Riemann curvature satisfies the Bianchi differential identity, vanishing of the Bianchi derivative, $R_{ab[cd_;e]}=0$.  Let's take its trace which leads to 
\begin{equation}
G^{ab}{}{}_{;b} = 0
\end{equation}
where
\begin{equation}
G_{ab} = R_{ab} - \frac{1}{2} R g_{ab},
\end{equation}
is the second rank symmetric tensor with vanishing divergence. Then we can write 
\begin{equation}
G_{ab} = -\kappa T_{ab} - \Lambda g_{ab}
\end{equation}  
with $T^{ab}{}{}_{;b}=0$ and the second term on the right is constant relative to covariant derivative. Could this be an equation of motion for the universal force responsible for inhomogeneity of spacetime curvature? On the left is a second rank symmetric tensor derived from Riemann tensor involving second order derivative of the metric and hence it is a second order differential operator like $\nabla^2$ operating on the metric potential $g_{ab}$. If we identify the new tensor $T_{ab}$ with energy-momentum distribution, which is universal, as source, then the above equation becomes equation of motion for the Einstein gravity. Thus emerges GR from spacetime curvature all by itself.  \\

The principle of Equivalence had played very important role in discovery of GR but we  made no reference to it in our derivation of the Einstein equation. This is simply because gravitational law being described by curved spacetime which admits a tangent 
plane at every point. This property of curved space automatically incorporates the Principle of equivalence. Since Einsteinian gravity is universal, this is why its equation of motion is geometric and so is motion under gravity -- a geodesic with no reference to any particle parameter. It is purely a geometric statement. The point to be noted is that Newton's second law does not apply to relativistic gravity because it is universal. For a geodesic motion there is no inertial and gravitational mass, what we need to experimentally verify is that how accurately particles follow the geodesic. Thus the question why should inertial and gravitational mass be equal becomes impertinent. \\
 
Note that we began by characterizing force or dynamics free state of spacetime and it is defined by (homogeneous) constant curvature. What happens when curvature is inhomogeneous,  the Einstein gravity naturally arises even though we had not asked for it. Like $c$ and $\Lambda$ characterize homogeneous spacetime, similarly the Einstein gravity characterizes inhomogeneous spacetime. In other words, gravity is inherent in inhomogeneity of spacetime curvature. This is different from rest of physics where a force law like Newton's gravity is always prescribed from outside. Thus the Einstein gravitational law cannot be prescribed instead it is dictated by inhomogeneity of spacetime. This is so simply because it is universal -- links to everything that physically exists, the unique distinguishing feature of the Einstein gravity. \\

Further note that $\Lambda$ enters into the equation on the same footing as energy-momentum tensor $T_{ab}$. It is therefore as solid a piece in the equation as energy-momentum, and hence it should not be subjected to one's whims and fancy without due physical justification. When $T_{ab}=0$, we are back to homogeneous spacetime of constant curvature. \\

\begin{svgraybox}
Had Einstein followed this line of reasoning to get to his equation, he would have certainly not treated $\Lambda$ as a blunder instead would have respectfully recognized it as a true constant of spacetime structure alongside the velocity of light. This would have saved us all from this monumental confusion that has gone over a century and yet no sign of diminishing. Further, perhaps he would have made the greatest prediction of all times that the Universe would experience accelerated expansion some time in the future. Had that been the case it would have been the most remarkable and truly Einstein like. Alas  that didn't happen.\\
\end{svgraybox}

The picture thus emerges is that homogeneity, characterizing force-free state, of spacetime requires two invariants, a velocity that binds space and time into spacetime and a length that gives a constant curvature to it. Introduction of matter/forces makes spacetime inhomogeneous and so emerges the Einsteinian gravitational dynamics. In the conventional picture, for absence of matter spacetime is taken to be flat and then matter makes zero curvature to non-zero. There is a discontinuity and break from flat to non-flat while in our picture there is continuous transition from homogeneity to inhomogeneity. There is therefore a paradigm shift, all dynamics free spacetime is thus not flat but is (homogeneous) of constant curvature and introduction of matter makes it inhomogeneous. As invariant velocity is needed to bind space and time into spacetime so as to provide a relativistic platform for physical phenomena, exactly with the same force of argument and spirit, $\Lambda$ is needed to provide an appropriate 
curved 
spacetime  platform for the relativistic gravitational phenomenon -- the Einstein gravity to unfold \cite{measure-grav}. \\

What it tells is that like constant potential is irrelevant for classical physics, similarly constant curvature is irrelevant for gravitational dynamics. It is because the  constant curvature spacetime is maximally symmetric characterizing the 'force free' state of spacetime. On the other hand, it turns out that constant potential for radially symmetric field in the usual Schwarzschild coordinates is indeed, unlike the Newtonian gravity, non-trivial for the Einstein gravity because it produces inhomogeneous spacetime of non-zero curvature \cite{b-v, ein-new}.

\section{Self interaction and vacuum energy} 

The driving force for GR is to universalize gravity which meant all forms of energy distribution including its own  self energy as well as zero mass particles must participate in gravitational interaction \cite{universal}. The only way zero mass particle can, since its velocity cannot change, be brought in the fold is that gravity must curve space and zero mass particle simply floats freely on it. Since space is already bound with time by the  invariant velocity, gravity thus curves spacetime. We have seen above how Einstein  gravity naturally follows purely from differential geometric property of Riemann curvature of spacetime. This is all very fine, but how is self interaction taken care of;i.e. how does gravitational field energy gravitate? Does it do through a stress tensor like any other matter field? No, we write no stress tensor on the right of the Einstein  equation given above. As a matter of fact gravitational potential in the Schwarzschild solution describing field of a mass point is the same 
Newtonian going as $1/r$ indicating that it is solution of the good old Laplace equation. There is no self interaction  contribution in it, and that is why the inverse square law remains intact. \\

If the self interaction were to be incorporated in the equation, then it should have been modified to $\nabla^2\Phi= 1/2 \Phi'^2$. This would have of course been relative to flat spacetime, and it would have modified the inverse square law. The latter is, as we all know, the cornerstone of classical physics as it ensures conservation of flux and thereby of charge. Hence that should not be tampered. The self interaction should therefore have to be accommodated without modifying the gravitational law. \\

Also note that the only way photon can respond to gravity is that gravity curves space. Could it be that the self interaction is responsible for curving space while the matter/energy distribution produces the inverse square law? This is exactly what happens in GR, and so we can say Einstein is Newton with space curved \cite{ein-new}. It is most remarkable that the two new aspects of GR, self interaction and photon feeling gravity, take care of each other so beautifully that the former curves space and that is precisely what is required for the latter. This is indeed the mark of sheer elegance and profoundity. \\

Why the self interaction is not visible in the Schwarzshild solution because it has been automatically absorbed in the space curvature when we write $g_{rr}=(1+2\Phi)^{-1}$ while the Newtonian acceleration is accounted for by $g_{tt}=1+2\Phi$. For solving the Einstein vacuum equation, when we write $R^t_t=R^r_r$ which demands $g_{tt}g_{rr}=-1$ and then  $R^t_t = \nabla^2\Phi=0$ leading to the inverse square law. This is how the equation requires space to be curved and the self interaction gets absorbed in that. If space were flat,  $g_{rr}=-1$, then $R^t_t=\nabla^2\Phi = {\Phi'}^2$, clearly showing the self interaction. When $g_{tt}=-g_{rr}^{-1}=1+2\Phi$, it gets beautifully absorbed in the space curvature leaving the Newtonian law intact \cite{ein-new}. \\

There is yet another subtlety that the Einstein potential can be zero only at infinity and nowhere else -- it is determined absolutely. This follows from $R^\theta_\theta=0$ which determines $\Phi=-M/r$ exactly without a possibility of addition of a constant. This happens because gravitational field energy can vanish only at infinity, and hence so should the space curvature. That means potential can only vanish at infinity. \\

The important lesson that follows is that gravitational field energy gravitates not through a stress tensor as a source on the right hand side of the equation but instead by enlarging the framework from flat to curved space. Why does this happen, what is it that is different for gravitational field energy? The answer is, that it is a secondary source which is produced by the primary source, matter-energy. It has no independent existence of its own -- it is matter fields that produce gravitational field. It is therefore natural that a secondary source produced by the primary source should not sit alongside in the equation. It can therefore only be incorporated by enlarging the framework \cite {measure-grav, cern}. \\

This suggests a general principle that anything that doesn't have independent existence of its own is a secondary source and hence \emph{must not} gravitate via a stress tensor but instead by enlarging the framework. The point in question is that of vacuum energy produced by quantum fluctuations of vacuum by the matter fields. It is exactly on the same footing as gravitational field energy. Never mind one is able to compute its stress tensor relative to flat spacetime, which has exactly the same form as $\Lambda g_{ab}$, it must not sit alongside the matter fields, $T_{ab}$. This is precisely the reason for its association with $\Lambda$ and then its incredible mismatch, $10^{120}$ with the Planck length. This is the root cause of the confusion which arises from making vacuum energy gravitate through a stress tensor. This defies and violates the above general principle just enunciated.  \\

If we adhere to the principle, there is no relation between $\Lambda$ and vacuum energy. It is then free to have any value that observations determine. Recall that both $c$ and $\Lambda$ arose purely from the symmetries of (homogeneous) spacetime as constants of its  structure. Then $c$ got identified with velocity of light and $\Lambda$ remained dangling until the 1997 supernova observations of accelerating Universe \cite{perl}. Thus $\Lambda$ simply represents acceleration of the Universe as all observations are wonderfully consistent with it. There is no need for any kind of dark energy involving exotic matter or outlandish modifications of gravitation theory. \\

Of course the moot question remains, how to enlarge the framework to make vacuum energy gravitate? Vacuum energy is a quantum creature and hence it would be difficult to guess the enlargement of framework until there emerges a quantum theory of gravity. If we had asked the same question for gravitational field energy in 1912, say, before the advent of GR, it would have been hard to guess that enlargement is in curving space. This means we won't know exactly without quantum gravity how to enlarge framework for making vacuum energy gravitate \cite{measure-grav, cern}. \\

There could however be some informed guesses based on the lessons learnt from  gravitational field energy. In GR, the real question was how to make light feel gravity which required space to be curved. Since space and time were already bound into spacetime by the velocity of light, it meant curving of spacetime. That is how gravity can only be described by spacetime curvature which automatically incorporated gravitational self interaction. Since the Newtonian inverse square law remains intact, self interaction can only curve space. This suggests that framework should be so enlarged that keeps GR intact. The real question therefore is to identify some phenomenon which has so far remained aloof, like light in the case of GR, and that has to be brought into the gravitational fold. Answering this question would require framework enlargement which would automatically incorporate gravitational interaction of vacuum energy. This is what will perhaps show the road to quantum gravity. Unfortunately we have not yet 
been able to clearly identify this critical question. That is the problem.  \\ 

Spacetime curves or bends like a material object, it should therefore have physical structure -- a micro-structure as is the case for any material object. That means space should have some micro building blocks -- "atoms of space". Such a micro structure is also required for vacuum to quantum fluctuate giving rise to vacuum energy. Thus micro structure of space is intimately related to vacuum energy and hence incorporation of the former would perhaps automatically, as anticipated, take care of gravitational interaction of the latter. The key question is then how to bring in atoms of space into the fray. Loop quantum gravity seems to follow this route but has not been able go far enough.  \\

Another possible avenue could be that vacuum energy may gravitate via higher dimension \cite{burgess, sandi, sds} leaving GR intact in the four dimensional spacetime. It is conceivable that at very high energy gravity may not entirely remain confined to four dimension, it may leak into higher dimension \cite{rand-sund}. The basic variable for gravity is the Riemann curvature tensor, for high energy exploration, we should include its higher powers in the action. Yet we want the equation of motion to retain its second order character, then this requirement  uniquely identifies Lovelock Lagrangian. Even though Lovelock action is a homogeneous polynomial of degree $N$ in the Riemann curvature, it has remarkable unique property that the resulting equation is always second order. Note that Lovelock gravity includes GR for $N=1$, and $N=2$ is the quadratic Gauss-Bonnet (GB) gravity, and then cubic and so on. But the higher order terms make non-zero contribution in the equation only in dimensions higher than four. 
If we want to explore high energy sector of gravity, which should indeed be the case for quantum gravity, we have to go to higher dimensions \cite{higher-d}. This is purely a classical motivation for higher dimensions. What it suggests is that the road to quantum gravity may go via higher dimensions notwithstanding the fact that higher dimensions are natural playground for string theory. Though string theory is a very popular approach to quantum gravity, yet it has also not gone far enough. \\

What it all suggests is that like quantum gravity, gravitational interaction of vacuum energy is an open question, and the solution of the latter is perhaps inseparable from the former. In the absence this, incorporation of it through a stress tensor is simply a tentative attempt similar to inclusion of gravitational self interaction by writing $\nabla^2\Phi= 1/2 \Phi'^2$. This was not borne out by the correct theory of gravity -- GR. So would be the case for vacuum energy when quantum gravity emerges. 

\section{In higher dimension} 

In the previous section, we have hinted that consideration of gravity in higher dimensions may not be entirely outrageous. Then the question arises, what should be the equation in there? Could it very well be the Einstein equation which is valid in all dimensions larger than two? Yes, that could be the case. However how did we land in four rather than three dimension? This is because in three dimension, gravity is kinematic which means Riemann is entirely determined by Ricci tensor and hence there exists no non-trivial vacuum solution. This translates into the fact that there are no free degrees of freedom for free propagation of gravitational field. This is how we come to four dimension where Riemann has 20 while Ricci has 10 components allowing for non-trivial vacuum black hole solutions. Could this feature be universalized for all odd dimensions in a new theory which reduces to Einstein gravity for dimension, $d\leq4$? It would be nice to incorporate this feature in higher dimension. \\

Another desirable feature that one can ask for is existence of bound orbit around a static object like a black hole. It is easy to see that in GR, bound orbits can exist only in $d=4$ and in none else. This is because gravitational potential goes as $1/r^{d-3}$ which becomes sharper and sharper with dimension while centrifugal potential always falls off as $1/r^2$ and hence the two can balance only in four dimension to give bound orbits. \\

If we take these two as the guiding features for gravitational equation in higher dimension, then pure Lovelock gravity is uniquely singled out \cite{dgj, eqn, dgj-bound, discern}. In Lovelock gravity, Lagrangian is $\sum\alpha_N{\cal L}^N$ where each $\alpha_i$ is a dimensionful coupling constant. Note that $\alpha_0=\Lambda$ is the cosmological constant and $\alpha_1=G,  \cal{L}=R$ are respectively the Newtonian gravitational constant and the Einstein-Hilbert Lagrangian. By pure Lovelock we mean that Lagrangian has the only one $N$th order term without sum over lower orders, and consequently the equation also has only one term. For pure Lovelock, gravitational potential goes as $1/r^{(d-2N-1)/N}$ for $d\geq2N+1$ \cite{dpp-sol}. For existence of bound orbits, what is required is $(d-2N-1)/N < 2$ which is always true for $N\geq1$, and it means $d<4N+1$. Further $d>2N+1$ else gravitational potential becomes constant, and hence we have the dimensional range, $2N+1 < d <4N+1$ for existence of bound orbits. For 
the linear $N=1$ Einstein, it is $3<d<5$ (and hence only $d=4$) while for the quadratic $N=2$ GB, $5<d<9$. \\ 

In pure Lovelock gravity, potential becomes constant in all critical odd $d=2N+1$ dimensions and hence gravity must be kinematic. For $N=1$ Einstein gravity, potential is constant in $d=3$, and gravity is kinematic in the sense that Riemann is given in terms of Ricci tensor. This means that it should be possible to define an $N$th order Lovelock Riemann tensor which is then given in terms of the corresponding Ricci in all $d=2N+1$ dimensions. This is indeed the case \cite{bianchi, kastor, xd}. The Lovelock Riemann, which is a homogeneous polynomial in Riemann, is defined by the property that vanishing of trace of its Bianchi derivative gives a divergence free second rank symmetric tensor -- Lovelock  analogue of Einstein tensor, and it is exactly the same as what one obtains by varying $N$th order Lovelock action \cite{bianchi}. Then the pure Lovelock gravitational equation reads as follows:

\begin{equation}\label{pure_Eq_01}
{}^{(N)}E^{a}_{b}\equiv -\frac{1}{2^{N+1}}\delta ^{ac_{1}d_{1}\ldots c_{N}d_{N}}_{ba_{1}b_{1}\ldots a_{N}b_{N}}R^{a_{1}b_{1}}_{c_{1}d_{1}}\ldots R^{a_{N}b_{N}}_{c_{N}d_{N}}= -8\pi T^{a}_{b}.
\end{equation}

It is then shown that $N$th order Lovelock Riemann can be entirely written in terms of  $N$th order Einstein tensor, ${}^{(N)}E^{a}_{b}$ \cite{xd}. The pure Lovelock gravity is  kinematic in all critical odd $d=2N+1$ dimensions. \\

We have identified the two critical properties of Einstein gravity, kinematicity in odd three dimension and existence of bound orbits around a static source, which we would like to carry over to higher dimensions. It is the universalization of these properties that leads to pure Lovelock equation uniquely. This is the right equation in higher dimensions \cite{eqn, discern}. For a given $N$, existence of bound orbits prescribes the  dimensional range, $2N+1 < d < 4N+1$. On the other hand stability of static black hole requires $d\geq3N+1$ \cite{c-d} and hence the range gets further refined to $3N+1 \leq d < 4N+1$. For $N=1$, it admits only one $d=4$ while for $N=2$, there are two $d=6, 7$, and in general number of allowed dimensions are equal to Lovelock order $N$. It is interesting that stability threshold is though included but not the entire range $d\geq3N+1$. That is, bound orbits exist for unstable black hole for $2N+1 < d < 3N+1$ while for $d \geq 4N+1$, black hole is stable without bound orbits around 
it. \\

Further pure Lovelock gravity gives rise to an interesting situation that $1/r$ potential on which whole of astrophysics and cosmology rest could occur not only in four but higher dimensions as well \cite{c-d}. This is because potential goes as  $1/r^{(d-2N-1)/N}$ which will be $1/r$ in all dimensions, $d=3N+1$. Static black holes are thus  indistinguishable in this entire dimensional spectrum. In particular four dimensional Schwarzschild black hole is indistinguishable from its pure GB seven dimensional counterpart. Not only that cosmology is also the same as FRW  expanding Universe evolves with the same scale factor \cite{c-d}. \\

So far as gravity is concerned, the situation is indistinguishable in dimensional spectrum $d=3N+1$ for all astrophysical and cosmological observations, except for gravitational degrees of freedom determining number of polarizations of gravitational wave. Number of degrees of freedom are given by $d(d-3)/2$ \cite{dof} which is two in four and 14 in  seven dimension. The Hulse-Taylor pulsar observations do verify two polarizations for emitted  gravitational wave. But for that it would not be possible to decide whether it is Einstein gravity in four or $N$th order pure Lovelock in $(3N+1)$ dimension. It is an  interesting feature of pure Lovelock gravity. 

\section{Outlook and perspective} 

GR is purely a principle and concept driven theory and it is therefore born as a whole 
complete theory. There was no observation or phenomenon driving it. That is why it was not developed as step by step but it emerged as a complete full 
theory. One can envisage the driving principle as inclusion of zero mass particle 
in mechanics and gravitational interaction. This meant universalization of mechanics and 
gravity for all particles including zero mass particles -- photons/light 
\cite{universal}. The former leads to relativistic mechanics known as special relativity while the latter to relativistic theory of gravitation -- general relativity. Of course in the former case there was the compelling phenomenological demand -- velocity of light was observed to be constant for all observers while for the latter there was no such phenomena asking for it. As a matter of fact, the first serious challenge to the Newtonian gravity only came as late as in mid 1960s in the form of observation of highly energetic quasi-stellar objects -- quasars. \\ 

It was a theory at least 50 years ahead of its times. This was because it was principle rather than experiment or observation driven. The same situation still holds as there is no strong observational challenge to it as yet. The accelerating Universe observation did pose some concern, and it did generate enormous amount of activity in building models of dark energy which were rather too many for comfort, and involved exotic matter fields and  outlandish modifications of the theory. However it has all settled down to $\Lambda$ successfully accounting for the observations. It is the symmetry of homogeneous spacetime that gives rise to $\Lambda$ as a true constant of spacetime structure on the same footing as velocity of light, and the accelerating Universe determines its value \cite{perl}. \\

Had Einstein followed the natural and straightforward geometric path to arrive at GR, he could have in fact realized the true significance of $\Lambda$ and would have predicted that the Universe would experience accelerated expansion some time in the future. That would have been the greatest prediction of all times. Then there won't have been any reason for questioning $\Lambda$ but instead one would have questioned how does vacuum energy stress tensor has the same form as $\Lambda$? Neither there would have been that much thrust for dark energy models nor would there have been acrobatics for getting $\Lambda$ as a constant of integration via trace-free or unimodular gravity \cite{ellis,paddy}. Though not much material difference but it would have been a different and perhaps the right perception. \\   

With this backdrop, the viewpoint, that vacuum energy cannot gravitate via a stress tensor but instead it would require an enlargement of framework, would have perhaps been appreciated with a positive disposition. It is then a principle that dictates that  secondary gravitational sources like self interaction and vacuum energy caused by  primary matter source do not gravitate via a stress tensor but instead they do by enlargement of the framework as is the case for the former -- by curving space \cite{measure-grav, cern}. For inclusion of vacuum energy in gravitational interaction, it is the principle that directs us to go beyond GR. It is therefore not only GR was principle driven but a journey beyond it is as well. Since vacuum energy is a quantum effect, for its inclusion we need a quantum theory of gravity. As and when it comes about, what is expected is that vacuum energy would be automatically included. The gravitational interaction of vacuum energy cannot be accommodated in GR itself and would 
remain an open question until quantum gravity is discovered.   \\ 

Another point to note is that the vacuum energy has equation of state, $\rho+p=0$, which defines inertial density for fluid equation of motion. What happens when inertial mass of a particle is zero, it cannot be accommodated in the existing framework. It asks for a  new framework of relativistic mechanics -- SR. Thus vanishing of inertial quantity is a serious matter, it indicates that it cannot be accommodated in the existing theory, and a new new theory would be required for its incorporation. \\ 
 
Once $\Lambda$ is liberated from vacuum energy, it can hence have any value that the observations determine. Then there is no embarrassing discrepancy of $10^{120}$ orders with the Planck length. What this number then indicates is simply the fact that the Universe measures this much in units of the Planck area \cite{cern}!  \\    

The important point to be noted is that universal velocity and universal force cannot be described by Newton's second law \cite{nd} instead they could only be described by spacetime geometry. Motion under gravity is free of particle mass, it simply follows geometry of spacetime -- geodesic. The response to gravity is therefore not through gravitational mass and hence passive gravitational mass is not defined. The question of equality of inertial and passive gravitational mass is therefore not admitted for the relativistic gravitational force. Therefore there is no need to pose the question of its equality with inertial mass, and so the formidable question of  explaining their equality doesn't arise. It is simply the reflection of the fact that gravity is no longer an external force, it is synthesized in spacetime geometry and hence Newton's second law is inapplicable. \\

Let me point out that light cannot bend, what bends is space. We measure bending of space by light because it freely floats on space \cite{subtle}. The question is how does space bend like a wire? Wire bends because it is made of small discrete units like atoms and molecules. This means discrete micro structure is necessary for anything to bend \cite{measure-grav}. At deep down Space should therefore also have discrete micro structure -- some fundamental entities as its building blocks. Given that, the question arises what is the natural geometric structure of such a system? Should it be flat with zero curvature or be of constant curvature. There are efforts afoot on building spacetime from an evolving system of causal sets. It turns out that constant curvature is more probable than zero curvature though the latter cannot be ruled out \cite{causalsets}. \\ 

For a classical force which is produced by a charge, when all charges in the Universe are summed over, the total charge must vanish. This is quite clear for electric charge because positive and negative charges are created by pulling out a positive or negative charge from a neutral entity, then what remains behind has opposite polarity. This must also be true for gravity. Energy-momentum is charge for gravity which is unipolar. How could it be balanced to yield total charge zero. The only possible way is that gravitational field must have charge of opposite polarity  \cite{universal}. This implies three things: gravity is self-interactive, negative gravitational charge is non-localizable and spread over whole space and gravity is always attractive. \\  

There is yet another general principle we would like to invoke is that all universal concepts must be related \cite{universal} and that relation must also be universal -- the same for all. By universality we mean a concept or phenomenon which is the same for all and equally shared by all. Space and time are universal entities and they must be related by a universal relation -- universal velocity. It binds space and time into spacetime. This then leads to the special relativity. Next gravity is a universal force, and hence its dynamics should be described by spacetime curvature -- general relativity \cite{why}. Is there anything else which is universal that could be bound to spacetime structure? Like gravity, the primary quantum uncertainty principle is also universal and hence it must also be related to spacetime. This is what has not been achieved and until that happens quantum theory remains incomplete. As and when that happens, it would give rise to a quantum theory of spacetime as well as of gravity. 
This is perhaps the 
deepest question of all times, probing the building blocks of spacetime itself. It is therefore the most formidable problem and it is not for nothing that it has so far defied all attempts by the best of the minds for over half a century. \\    

Of late there has been lot of work on gravity as an emergent force. It began with the seminal work of Ted Jacobson who could deduce the first law of thermodynamics from the Einstein vacuum equation \cite{ted}. The activity picked up considerably in past decade or say, and among others, Paddy along with his coworkers is one of the leading players \cite{paddy-review}. It raises the question, if it is emergent like thermodynamical laws, it is not a fundamental force. It is a bulk property of some underlying kinetic structure of "atoms of space". It is indeed a very deep question, is gravity fundamental or emergent? I believe that the question is similar to asking, is photon a wave or particle?  Both gravity and photon are self-dual, meaning their dual is contained in themselves. Gravity is different from all other forces in several ways, and perhaps the most remarkable of them all is that it is both fundamental and emergent or neither. \\

Notwithstanding all this, probing and understanding of quantum structure of building blocks of spacetime is very pertinent and is the most challenging question of the day. \\ 

Finally I conclude it with cheering Paddy warmly and affectionately on turning 60, which is simply a number like any other, and it doesn't matter at all if one doesn't mind it. It is all anyway a matter of mind.


\begin{thebibliography}{90}

\bibitem{measure-grav} N. Dadhich, Int. J. Mod.Phys. {\bf D20}, 2739 (2011), arxiv: 1105.3396.

\bibitem{ein-new} N. Dadhich, Current Science {\bf 109}, 260 (2015); arxiv:12060635.

\bibitem{b-v} M. Barriola and A. Vilenkin, Phys. Rev. Lett. {\bf 63}, 341 (1989).

\bibitem{universal} N. Dadhich, "Universalization as a physical guiding principle", gr-qc/0311028.

\bibitem{perl} S. Perlmutter et al, Astrophys. J. {\bf 483}, 566 (1997); Nature {\bf 391}, 51 (1998).

\bibitem{cern} N. Dadhich, "$\Lambda$ is a constant of spcetime structure and has nothing to do with vacuum energy", The talk given in the Workshop on Questioning Fundamental Physical Principles, CERN, May 6-9, 2014 and in MG14, Rome, 2015.

\bibitem{burgess} C. P. Burgess, Extra dimensions and the cosmological constant problem, arxiv:0708.0911. 

\bibitem{sandi} Sandipan Sengupta, Class. Quant. Grav. {\bf 32}, 195005 (2015); arxiv:1501.0079.

\bibitem{sds} V. Soni, N. Dadhich, R. Adhikari, The cosmological constant from the zero point energy of compact dimensions; arxiv:1509.02156.

\bibitem{rand-sund} L. Randall and R. Sundrum, Phys. Rev. Lett. 83, 4690 (1999).

\bibitem{higher-d} N. Dadhich, "Universalization as a physical guiding principle", gr-
qc/0311028; "Universal features of gravity and higher dimensions", arxiv:1105.0988; "Universality, gravity, the enigmatic $\Lambda$  and beyond", Proceedings of 12th Regional Conference on Mathematical Physics, eds. M. J. Islam, F. Hussain, A. Qadir, H. Saleem, (World Scientific, 2006), hep-th/0509126.
 
\bibitem{dgj} N. Dadhich, S. G. Ghosh, S. Jhingan, Phy. Lett. {\bf B711}, 196 (2012), arxiv:1202.4575.

\bibitem{eqn} N. Dadhich, "The gravitational equation in higher dimensions", in Relativity and Gravitation: 100 years after Einstein in Prague, eds J. Bicak, T. Ledvinka (Springer, 2013), arxiv:1210.3022.

\bibitem{dgj-bound} N. Dadhich, S.G. Ghosh, S. Jhingan, Phys. Rev. {\bf D88}, 124040 (2013), arXiv:1308.4770.

\bibitem{discern} N. Dadhich, Euro. Phys. J. {\bf D 78}, 1 (2016); arxiv:1506.08764.

\bibitem{dpp-sol} N. Dadhich, J. M. Pons, K. Prabhu, Gen. Rel. Grav. {\bf 45}, 1131 (2013), arxiv:1201.4994.

\bibitem{bianchi} N. Dadhich, Pramana {\bf 74}, 875 (2010), arxiv:0802.3034.

\bibitem{kastor} D. Kastor, Class. Quant. Grav. {\bf 29}, 155007 (2012), arxiv:1202.5287.

\bibitem{xd} X. O. Camanho, N. Dadhich, On Lovelock analogues of the Riemann tensor, arxiv:1503.02889.

\bibitem{c-d} S. Chakraborty, N Dadhich, Do we really live in four or higher dimension; arxiv:1605.01961.

\bibitem{dof} N. Dadhich, R. Durka, N. Merino, O. Miskpvic, Phys. Rev. {\bf D 93}, 064009 (2016); arxiv:1511.02541. 

\bibitem{ellis} G. Ellis, Henk van Elst, J. Murgan and J.-P. Uzan, Class. Quant. Grav. {\bf 28}, 225007 (2011); arxiv:1008.1196.

\bibitem{paddy} T. Padmanabhan and H. Padmanabhan, Int. J. Mod. Phys. {\bf D23}, 1430011 (2014); arxiv:1404.2284.

\bibitem{nd} N. Dadhich, Physics News {\bf 39}, 20 (2009); arxiv:1003.2359. 

\bibitem{subtle} N. Dadhich, Subtle is the gravity, gr-qc/0102009. 

\bibitem{causalsets} M. Ahmed, D. Rideout, Phys. Rev. {\bf D 81}, 083528 (2010); arxiv:0909.4771. 

\bibitem{why} N. Dadhich, Why Einstein (Had I been born in 1844!)?; physics/0505090.

\bibitem{ted} T. Jacobson, Phys. Rev. Lett. {\bf 75}, 1260 (1995). 

\bibitem{paddy-review} T. Padmanabhan, Exploring the nature of gravity, arxiv:1602.01474; S. Chakraborty and T. Padmanabhan, Phys. Rev. D 92, 104011 (2015) arXiv:1508.04060; T. Padmanabhan, S. Chakraborty and D. Kothawala, Gen. Relt. Grav. 48, 55 (2016) arXiv:1507.05669.



 


























\end{thebibliography}
\end{document}